\documentclass{article}

\usepackage{setspace}
\usepackage{macros/packages}
\usepackage{macros/statistics}
\usepackage{macros/editing}
\usepackage{macros/formatting}
\hypersetup{colorlinks,citecolor=blue,urlcolor=blue,linkcolor=blue}
\graphicspath{ {./images/} }
\usepackage{multicol}
\usepackage{graphics}
\usepackage{nicematrix}
\definecolor{Gray}{gray}{0.9}
\definecolor{bg}{rgb}{0.95, 0.95, 0.96}

\usepackage{xr}
\makeatletter
\newcommand*{\addFileDependency}[1]{
  \typeout{(#1)}
  \@addtofilelist{#1}
  \IfFileExists{#1}{}{\typeout{No file #1.}}
}
\makeatother

\newcommand*{\myexternaldocument}[1]{
    \externaldocument{#1}
    \addFileDependency{#1.tex}
    \addFileDependency{#1.aux}
}

\myexternaldocument{supplement}
\begin{document}

\textbf{Title}: Clinical Utility Gains from Incorporating Comorbidity and Geographic Location Information into Risk Estimation Equations for Atherosclerotic Cardiovascular Disease 

\vspace{0.25in}

\textbf{Corresponding author}: Yizhe Xu\\ 
\textbf{E-mail}: yizhex@stanford.edu \\
\textbf{Postal address}: 3180 Porter Dr., Palo Alto, CA, 94304\\
\textbf{Telephone}: (801) 433-7346

\vspace{0.25in}

\textbf{Co-authors}: Yizhe Xu, Agata Foryciarz, Ethan Steinberg, and Nigam H. Shah\\
\textbf{Affiliation}: (shared by all co-authors) Stanford Center for Biomedical Informatics Research, Stanford University, Stanford, CA.

\vspace{0.25in}


\vspace{0.25in}
\textbf{Word count}: 6,000

\newpage

\begin{abstract}
\textbf{Objective:} There are several efforts to re-learn the 2013 ACC/AHA pooled cohort equations (PCE) for patients with specific comorbidities and geographic locations. With over 363 customized risk models in the literature, we aim to evaluate such revised models to determine if the performance improvements translate to gains in clinical utility.

\textbf{Methods:} We re-train a baseline PCE using the ACC/AHA PCE variables and revise it to incorporate subject-level geographic location and comorbidity information. We apply fixed effects, random effects, and extreme gradient boosting models to handle the correlation and heterogeneity induced by locations. Models are trained using 2,464,522 claims records from Optum\textsuperscript{©}’s Clinformatics\textsuperscript{®} Data Mart and validated in the hold-out set (N=1,056,224). We evaluate models' performance overall and across subgroups defined by the presence or absence of chronic kidney disease (CKD) or rheumatoid arthritis (RA) and geographic locations. We evaluate models' expected net benefit using decision curve analysis and models' statistical properties using several discrimination and calibration metrics.

\textbf{Results:} The baseline PCE is miscalibrated overall, in patients with CKD or RA, and locations with small populations. Our revised models improved both the overall (GND P-value=0.41) and subgroup calibration but only enhanced net benefit in the underrepresented subgroups. The gains are larger in the subgroups with comorbidities and heterogeneous across geographic locations.  

\textbf{Conclusions:} Revising the PCE with comorbidity and location information significantly enhanced models' calibration; however, such improvements do not necessarily translate to clinical gains. Thus, we recommend future works to quantify the consequences from using risk calculators to guide clinical decisions.
\end{abstract}

\newpage

\section{Introduction}
\label{sect:intro}  
Atherosclerotic cardiovascular disease (ASCVD) and stroke are leading causes of deaths in the U.S. and cost healthcare systems about \$330 billion each year\citep{branchard2018glance}. To manage these conditions, clinicians often evaluate patients' risk of having a major adverse cardiovascular event using an equation and select a treatment strategy based on the estimated risk. So far, the 2013 ACC/AHA pooled cohort equations (PCE)\citep{pce13} is the most commonly used risk estimator to inform statin initiation, by following the 2019 ASCVD prevention guideline from the American College of Cardiology and the American Heart Association (ACC/AHA)\citep{arnett2019}. However, several studies have found that PCE is out of calibration, which leads to substantial misestimation of patients' ASCVD risk\citep{cook2016calibration, yadlowsky2018clinical, damen2019performance} and further exacerbates health inequalities\citep{foryciarz2022evaluating}. In addition, Rodriguez\citep{rodriguez2019atherosclerotic} and Khera\citep{khera2020performance} pointed out that PCE overestimates the ASCVD risk for non-hispanic whites, African Americans, Asians, Hispanics and subjects in the overweight and obese categories, respectively.

Given the impact of the PCE's risk estimates on ASCVD treatment decisions, a large number of efforts have been made to address this miscalibration by proposing revised models that utilize larger data sources\citep{wallisch2020re} or apply advanced machine learning approaches\citep{yadlowsky2018clinical} or enforce fairness constraints\citep{pfohl2019creating, barda2021addressing}. Such efforts to revise the PCE may very well double the 363 ASCVD risk prediction models that already existed by 2013\citep{damen2016prediction}. A series of systematic reviews have recently been conducted to summarize the common risk scores (e.g., FRS, QRISK, and SCORE) and modeling approaches\citep{faizal2021review}, predictive models in particular populations\citep{cai2020prediction}, and how social determinants of health are used\citep{zhao2021social} to help digest the large body of ASCVD risk estimation tools. Thus, it is time to pause and ask the question that whether the revised risk models lead to changes in clinical outcomes or resource allocation.

Typically, risk prediction models are evaluated in internal and external data sets using two metrics: discrimination and calibration. C-statistic is the most common discrimination measure that quantifies the level of concordance between the predicted risk given by a model and the observed risk\citep{harrell1996multivariable, pencina2004overall, gerds2013estimating}. In contrast to C-statistic, which summarizes how well the individuals risks are ranked, calibration quantifies how close the predicted risks are to observed risks\citep{demler2015tests}. The terms \emph{over-} or \emph{under-calibration} are used to describe situations where the estimated risks are larger or smaller than observed risks, respectively, which are both referred to as being \emph{out-of-calibration}. Although these two metrics are critical for evaluating the reliability of a risk model, a discrimination measure may depend on the risk variability in the dataset used for evaluation, and it is difficult to derive a model that achieves good calibration for all patient subgroups, such as those defined by geographic location or the presence/absence of certain comorbidities. We argue that, it is critical to also measure the potential gains from using a risk model to guide an intervention\citep{chohlas2021learning, shah2019making}. Therefore, metrics that consider clinical consequences, e.g., benefits and harms of an intervention, should be emphasized. For instance, the net benefit from decision-curve analysis\citep{vickers2006decision, Vickers2016} and the incremental cost-effectiveness ratio from cost-effectiveness analysis\citep{moons2012risk, pandya2015cost} are commonly-used mechanisms to quantify the consequences of model-guided decision making.  

In this work, we have three aims: First, we perform a comprehensive evaluation of the baseline PCE focusing on its clinical utility quantified using net benefit from decision curve analysis. We re-train the 2013 ACC/AHA PCE with a similar set of covariates using the large claims data from Optum\textsuperscript{©}’s Clinformatics\textsuperscript{®} Data Mart (CDM) ZIP5 \citep{Stanford_Center_for_Population_Health_Sciences_Optum_ZIP5_2021} then validate it internally in the held-out Optum ZIP5 data and and externally in the electronic health record data from Stanford medicine research data repository (STARR). We evaluate the overall net benefit, discrimination, and calibration of PCE, and the subgroup performance across different geographic locations and co-morbid conditions. Second, we revise the baseline PCE by including additional patient-level information on residence zip code and comorbidities. Local-area estimation of ASCVD risk helps to capture unmeasured factors such as air pollution, social determinants of heath, and neighbourhood influences on nutrition and physical activity. The widely-used QRISK3 risk calculator in the UK includes a deprivation score that enables such local estimates\citep{hippisley2017development}. Also, compared to QRISK3, the ACC/AHA PCE only includes the diabetes comorbidity variable, so we consider two other risk factors for ASCVD: chronic kidney disease (CKD) and rheumatoid arthritis (RA)\citep{jankowski2021cardiovascular, semb2020atherosclerotic}. We apply a fixed effects model and a random effects model to capture both the correlations and heterogeneity between location subgroups, and an extreme gradient boosting (XGB) model for estimating nonlinear relationships between covariates and outcome. Third, we convert the differences in net benefit between revised models and the baseline PCE to quantities such as the number of correct treatments and the number of unnecessary interventions avoided. Doing so summarizes the clinical implications of using the baseline PCE and revised risk models for assessing patients' ASCVD risk and informing treatment decisions. 


\section{Method} \label{sect:method}
\subsection{Data Sources and Formulation of the Study Cohort}
Optum CDM is a de-identified health claims database that contains over 51 million patients. This large database contains data on medical and pharmacy claims and lab results from 2003 to 2021. To evaluate the performance of adding geographic location information to the PCE equation, we utilize a variant of Optum data, the \emph{Optum ZIP} database version 5.0, that includes 5-digit zip code information\citep{Stanford_Center_for_Population_Health_Sciences_Optum_ZIP5_2021}. Approval for the usage of this data for this study was granted by the institutional review board of Stanford University (eProtocol \#IRB-46829).

We applied similar eligible criteria described in the current clinical practice guidelines for statin initiation\citep{pce13, arnett2019} to extract the primary study cohort from Optum ZIP. First, in order to obtain an analysis data set that captures the ACC/AHA PCE variables, we removed all patients without high-density lipoprotein (HDL) or cholesterol measurements. In addition, we restricted the cohorts to patients between 40 and 75 years old with no prior history of cardiovascular disease events or statin use at the time of visit. We also required patients to have HDL is between 20 and 100 mg/dL and total cholesterol is between 130 and 320 mg/dL. We attempted to also require systolic blood pressure (SBP) measurements, but were unable to find sufficient patients with that criteria (and instead replaced all SBP variables with binary indications of prior hypertension diagnoses). We refer to the final data set as the\emph{PCE-eligible} cohort. 

The STARR is a clinical warehouse that contains approximately 150 million encounter records from the Stanford Health Care and Lucile Packard Children's Hospital. This database comprises longitudinal data on patient and provider information, diagnoses, procedures, medication orders, and lab results that have been normalized to the Observational Medical Outcomes Partnership (OMOP) CDM version 5.3.1\citep{Datta2020}, sourced from inpatient and outpatient clinical encounters that occurred between 1990 and 2021. The use of this data was conducted in accordance with all relevant guidelines and regulations. Approval for the use of STARR for this study is granted by the Stanford Institutional Review Board Administrative Panel on Human Subjects in Medical Research (eProtocol \#IRB-46829), with a waiver of informed consent. 

In our study, the STARR data is used for two purposes: 1) To quantify the combined impact of using the hypertension indicator to replace the SBP variable and not using race as a predictor in the baseline PCE on its performance. This is because the Optum ZIP data does not contain information on race and there is a substantial amount of missing data on SBP, but these two variables are available in STARR; 2) To serve as an external validation data set for the ACC/AHA PCE and re-traind baseline PCE. To create the analysis data set, we followed the same algorithms used for Optum ZIP but including the continuous SBP variable and race instead. 

For both Optum ZIP and STARR, we require every subject to have at least one year of data prior to their index date, and we generate the final cohort by randomly sampling one record for each patient from the resulting candidate index events. The sample sizes of the PCE-eligible cohorts from Optum ZIP and STARR are 3,520,746 and 28,896, respectively. The distribution of baseline covariates are summarized in Table \ref{t:summary}.
\begin{table}
\centering
\caption{Summary of Baseline Characteristics in Optum ZIP and STARR data. Continuous variables are summarized as mean and standard deviation and categorical variables are summarized as count and percentage. The total sample size (including ineligible patients) of  are 16,792,476 and 262,161, respectively.}
\label{t:summary}
\begin{tabular}{lcc}
\hline
Variable & Optum ZIP (N = 3,520,746) & STARR (N = 28,896) \\ \hline
Age (years) &	53.5 (9.9) &	55.9 (9.4)\\
Female     &	2,116,013 (57.4\%)&	16,316 (56.5\%)\\
Current Smoker	& 328,486 (8.9\%)	& 24,449 (84.6\%)\\
High-density Lipoproten         &	55.8 (17.4)&	58.1 (16.4)\\
Total Cholesterol &	197.0 (37.1) &	195.0 (34.2)\\
Antihypertensives  & 1,081,759 (29.3\%) &	4,193 (14.5\%)\\
Diabetes &	499,243 (13.5\%)&	1,894 (6.6\%)\\
Hypertension &	1,439,459 (39.0\%)&	7,194 (24.7\%)\\
Chronic Kidney Disease      	& 	109,116 (3.0\%) &	640 (2.2\%)\\
Rheumatoid Arthritis 	        &	64,641 (1.8\%)&	232 (0.8\%)\\
Systolic Blood Pressure         &	--  &	126.4 (16.6)\\
\multicolumn{3}{l}{Race and Ethnicity}	\\
\,\,\,Black or African American &	-- &	2,168 (7.5\%)\\
\,\,\,White   & -- &	26,728 (92.5\%)\\
\hline
\end{tabular}
\end{table}

\subsection{Outcome and Feature Extraction}
We study the composite ASCVD outcome by estimating the 5-year risk of a fatal coronary heart disease, myocardial infarction, or stroke event. We consider coronary heart disease to be fatal if death occurs within a year after being diagnosed. The censoring occurs when a subject started using statin, died, or reached the end of the follow-up period (i.e., when subjects switch to a difference insurance for Optum ZIP; the last record in the Stanford health system for STARR). For both data sources above, we created the ASCVD event indicator according to these definitions and computed the follow-up time as the time from an index date to the date of having an event or being censored. 

We used OMOP CDM concept identifiers to extract time-agnostic demographic features (race and ethnicity, sex, and age), drug exposures (antihypertensives), lab test results (HDL, total cholesterol, SBP), and condition occurrences (diabetes mellitus, chronic kidney disease, rheumatoid arthritis, hypertension, and smoking history), and the 5-digit zip code. After searching using Logical Observation Identifiers Names and Codes (LONIC), we found little SBP information in Optum ZIP due to the fact that Optum started collecting data on SBP around 2019. Thus, we extracted the hypertension condition as a surrogate variable for SBP. In general, the lab results are not well recorded in claims and electronic health records databases, so we focus on the PCE-eligible cohort defined above instead of conducting multiple imputations as the assumption of missing at random may not hold in our situation. Important concepts and corresponding codes are presented in Table \ref{t:concept_code} using the Systematized Nomenclature of Medicine (SNOMED) concept codes. 

\subsection{Model Learning}\label{sec:model}
\subsubsection{The Baseline PCE}
2019 ACC/AHA guideline\citep{arnett2019} on ASCVD prevention recommended using the ACC/AHA PCE to assess the risk for individuals in varied age groups. The ACC/AHA PCE were derived using Cox proportional hazard (PH) models for four demographic groups defined by sex (female versus male) and race and ethnicity (African American versus White). Due to the limited SBP data in Optum ZIP, we re-train the ACC/AHA PCE using a similar set of covariates where SBP is replaced by a binary indicator of whether a subject has a hypertension condition at baseline. Moreover, we leave out the race variable due to the lack of race information in Optum ZIP and the arguments on using race for risk estimation may worsen racial inequity\citep{vyas2022revising}. Finally, we fit one Cox PH model with sex being one of the covariates instead of developing two separate models for male and female groups as in the ACC/AHA PCE, i.e., 
\begin{equation}
    \lambda_j(t) = \lambda_0(t)\,\mathrm{exp}(\beta^\intercal X_j), \label{eq:coxph}
\end{equation}
where $\lambda_0(t)$ and $\lambda_j(t)$ are the baseline hazard and the hazard of subject $j$ at time $t$, respectively, and $\beta$ is the vector of coefficients (effects on the log hazard scale) of the PCE variables. The five-year ASCVD risk is then computed as $1-S_j(t)$, where $S_j(t) = \mathrm{exp}(-H_j(t))$ with the cumulative hazard $H_j(t) = \int_{0}^{t}\lambda_j(u)\mathrm{d}u$ and $t = 5$ years. This re-trained model is used as the surrogate of the ACC/AHA PCE and referred to as \emph{baseline PCE}.

Furthermore, we perform a sensitivity analysis using the STARR data to quantify the difference between baseline PCE and ACC/AHA PCE. We apply both equations separately to estimate the five-year risk for all the PCE-eligible STARR subjects then compare their models' performance. As STARR is an external data source to the derivation data of ACC/AHA PCE and Optum ZIP, this assessment is an external validation. 

\subsubsection{The Revised PCE}
Given the consistent prior evidence on the ACC/AHA PCE being inaccurate, we revise the baseline PCE by utilizing additional subject-level information on comorbidities and geographic locations. CKD and RA are two risk-enhancing factors that are recommended to consider for adults at borderline (5 - 7.5\%) and intermediate risk (7.5 - 20\%)\citep{arnett2019}. CKD is typically determined by an estimated glomerular filtration rate value of 15-59 mL/min/1.73m\textsuperscript{2} with or without albuminuria and not treated with dialysis or kidney transplantation. RA is a systemic autoimmune inflammatory disease that affects many body parts including the circulation system and patients with RA have double the ASCVD risk\citep{semb2020atherosclerotic}. We include these two co-morbid conditions as binary variables in multivariate models. 

Individuals who resident in the same area may share common environmental (e.g. air pollution), lifestyle (e.g., diet and physical activity), and other social determinants of heath factors, so their survival outcomes may be positively dependent, i.e., the random errors that are absorbed into the baseline hazard $\lambda_0(t)$ are not independent and identically distributed (i.i.d.). In addition, the differences across local-areas may contribute to the unobserved heterogeneity, which may improve individual-level risk estimation when taken into consideration. Thus, we also incorporate individuals' zip code information to enable local-area risk estimation. To deal with the issue that some location subgroups have small populations, we group subjects using the code of sectional center facility (i.e., the first three digits of the 5-digit zip code) and merge a small group with the nearest location iteratively to ensure a minimum sample size of 3,000 within each location subgroup. To account for the correlation and heterogeneity issues induced by locations, we construct the following three revised models: First, a fixed effects model that includes the location variable as a factor, i.e., 
$$\lambda_{ij}(t) = \lambda_0(t)\,\mathrm{exp}(\gamma_i + \beta^\intercal X_{ij}),$$
where $i$ and $j$ are the location and individual indices, respectively, and $\gamma_i$ denotes the average (constant) effects for location $i$. $X_{ij}$ represents the baseline PCE covariates of subject $j$ in the location $i$ and $\beta$ is the vector of corresponding effect estimates. With a Cox regression, the first location is used as the reference level, i.e., $\gamma_1=0$, and the effects in other locations compared to the first location are the coefficients $\gamma_i$, $i\ne1$, by the model. 

Second, a random effects model by assuming the location subgroups are random draws from a larger population of locations, i.e., $$\lambda_{ij}(t) = Z_i\lambda_0(t)\,\mathrm{exp}(\beta^\intercal X_{ij}),$$ where $Z_i$ is the random location effect that follows a gamma distribution in our case, so it is also called a random effects model. In this model, we assume the random effect $Z$ is shared among all the subjects within the same location $i$, and it is constant over time. Compared to the fixed effects model, the random effects model shrinks the local estimates towards the mean estimates over all locations, which helps to improve inferences when the sample sizes of some location subgroups are small.  

Third, an extreme gradient boosting (XGB) model to estimate the ASCVD risk from a data-driven perspective\citep{chen2016xgboost}. XGB is a salable and fast implementation of the gradient boosting machine method that handles sparse data and instance weights in approximate tree learning. The detailed explanation of the XGB approach is given in Section \ref{XGB}. To obtain a top performing model, we tuned four hyperparameters, including the minimum number of observations in tree terminal nodes, the maximum depth of predictor interactions, the fractions of row and column subsampling, and the best set of hyperparameters (1000, 3, 0.9, 1, respectively) was chosen through grid search using five-fold cross-validation. As a slower learning rate often leads to better estimation performance but requires a large number of iterations, we used a learning rate of 0.1 with 100 trees (a.k.a iterations) during the hyperparameter tuning process and reduced it to 0.05 with 500 trees in the final model. The optimal number of boosting iterations was determined by applying early stopping, i.e., the model training is stopped if the negative partial log-likelihood in the validation set is not improved for 20 rounds. Our analyses using XGB were carried out by the R package \emph{xgboost}.

For both the baseline PCE and revised models, we randomly sampled 70\% of the Optum ZIP data as the training set and used the rest as the testing set for internal validation. For the XGB model, the hyperparameters were tuned by cross-validating the training data.

\subsection{Evaluation Metrics for Risk Models}
We evaluate the baseline PCE and the revised risk models for the purposes of benchmarking and quantifying the performance gains relative to baseline, respectively. For each model, we consider two evaluation schemes: \emph{Overall evaluation} and \emph{subgroup evaluation}. Overall evaluation is based on the risk estimates of subjects in the entire testing set, while subgroup evaluation is conducted for each subgroup at a time. We also assess how much the model performance vary across local-areas and comorbidity subgroups, which provides evidence on the robustness of these ASCVD risk estimators. We conduct the internal validation on the Optum testing data and the external validation on STARR data using the evaluation metrics described below.

\subsubsection{Statistical performance metrics}
For model discrimination, we measure the concordance (\emph{C}) statistic as the proportion of concordant pairs of subjects in which the concordance is defined as individuals with shorter survival times also have lower predicted survival probabilities, i.e., $\mathbb{P}(\ind{M_n(t, X_i)>M_n(t, X_j)} | T_i<T_j, T_i \leq t)$, where $T_i$ and $T_j$ are the survival times of the two subjects in a random pair and $M_n(\cdot)$ is a risk prediction model, and $t$ is the estimation horizon. Note that we consider a truncated version of the Harrell's C-statistic as the estimated survival time function is often unstable at tails\citep{harrell1996multivariable, pencina2004overall}. In the presence of right censoring with a censoring time $C_i$, we only observe the survival times for some subjects, i.e., $\Delta_i = \mathbb{1}\{T_i \leq C_i\}$, and the others are censored ($\Delta_i=0$). Let $\widetilde{T}_i=\mathrm{min}(T_i, C_i)$ be the observed follow-up time and $\widetilde{N}_i(t)=\mathbb{1}\{\widetilde{T}_i \leq t \,\, \mathrm{and} \,\, \Delta_i=1\}$, then the C-statistic is computed using a Mann-Whitney statistic for estimating the area under the Receiver Operating Curve:
$$\what{C}_\mathrm{HC}(t) = \frac{\frac{1}{m^2}\sum_{i=1}^{m}\sum_{j=1}^{m}\ind{M_n(t, X_i)>M_n(t, X_j)} \mathbb{1}\{\widetilde{T}_i < \widetilde{T}_j\} \widetilde{N}_i(t) }{\frac{1}{m^2}\sum_{i=1}^{m}\sum_{j=1}^{m}\ind{\widetilde{T}_i < \widetilde{T}_j} \widetilde{N}_i(t)},$$
where $m$ is the number of subjects. Note that this truncated C-statistic ignores the unusable pairs (the short survival time in a pair is censored), so it does not utilize all the information and its limiting value depends on the censoring distribution\citep{uno2011c}. We describe another discrimination metric that models censoring as a functions of covariates\citep{gerds2013estimating} in Section \ref{IPCW-C}, and the results from these two C-statistics are shown in Table \ref{t:cindex_comp}.

For calibration, the central idea is to measure how close the predicted risks are to the observed ones. Calibration-in-the-large is a basic metric that measures such agreement at the overall (mean) level; it is measured as the ratio of the number of observed to expected events, commonly denoted as O/E, with an ideal value of 1. One can compute the O/E value by counting the events in the data, but when a Cox PH risk model is applied, the O/E is also the intercept $\alpha$ in the following Poisson model\citep{crowson2016assessing}:
$$\mathrm{log}(\mathbb{E}[\delta|X]) = \alpha + 1\cdot R, $$
where $R = \mathrm{log}(H_i(\widetilde{T}_i)) = \mathrm{log}(H_0(\widetilde{T}_i)) + \beta^\intercal X_i$ and $H_i(\widetilde{T}_i)$ is the expected number of events when evaluated at the end of follow-up time $\widetilde{T}_i$; $H_i(t)$ is the cumulative hazard given in Eq (\ref{eq:coxph}). To measure O/E for subgroups, we create a group variable $K$ that may be defined by risk scores or other factors such as locations then include it into the Poisson model as a factor without a reference level, i.e., $\mathrm{log}(\mathbb{E}[\delta|X]) = \gamma^\intercal K + 1\cdot R$, where the length of the vector $\gamma$ equals to the number of groups. 

To examine calibration at a finer level, we apply the standard Greenwood–Nam–D’Agostino (GND) test to determine whether a model is miscalibrated\citep{demler2015tests}. GND test, as an extension of the Hosmer-Lemeshow test to survival outcomes, computes the mean \emph{observed-predicted} number of events within each of the $K$ risk groups (e.g., deciles), where \emph{observed} are computed using a KM estimator. The test statistic is distributed as a chi-square random variable with $K-1$ degrees of freedom, i.e., 
$$\chi_\mathrm{GND}^{2} (t) = \sum_{k=1}^{K}\frac{\left[\mathrm{KM}_k(t)-\overline{p_k(t)}\right]^2}{\mathrm{Var}(\mathrm{KM}_k(t))} \sim \chi_{K-1},$$        
where $\mathrm{KM}_k(t)$ and $\overline{p_k(t)}$ are the observed and predicted event probabilities and $\mathrm{Var}(\mathrm{KM}_k(t))$ is the Greenwood variance estimator of $\mathrm{KM}_k(t)$\citep{greenwood1926natural}. Although the GND test can deal with varied censoring levels and different types of prediction models, the results depend on the number of risk bins, the size of each bin, and the number of events in each bin. We choose the optimal number of bins $K = N^{1/3}$ suggested in \citet{nevat2021bin} that minimizes the mean squared errors of a Regressogram, where $N$ is the sample size of the evaluation set.

We also generate calibration plots from the GND test as the graphical assessments, in which the observed and expected risks are plotted for each bin and the 95\% confidence intervals (CIs) of the observed risks are computed using the Greenwood standard error above. Similar to the O/E calculation above, we extract the calibration intercept $\alpha$ and slope $\beta$ from the Poisson model below: 
$$\mathrm{log}(\mathbb{E}[\delta|X]) = \alpha + \beta^\intercal R, $$
which also provides 95\% CIs\citep{crowson2016assessing}. The ideal value of a calibration slope is 1 and of a calibration intercept is 0. 

\subsubsection{Clinical utility metrics}
Even though the discrimination and calibration evaluations have straightforward interpretations from a statistical perspective, they do not directly measure the clinical impact of using a risk model to guide interventions. To assess the effect on treatment recommendations, we conduct decision curve analysis and compare model performance using net benefit that considers the trade-off between benefits and harms of an intervention\citep{Vickers2016}. Net benefit quantifies the clinical utility of a model by incorporating the clinical consequence, specified by an exchange rate, into the calculation. For instance, if a doctor is willing to treat five subjects with statin in order to prevent one ASCVD event, then the exchange rate is 20\%. So, the exchange rate can be viewed as a threshold of ASCVD risk used to determine whether or not to initiate statin. Specifically, the net benefits (NB) are computed as
$$ \mathrm{NB} = \frac{|\mathrm{TP}|}{N} - \frac{|\mathrm{FP}|}{N} \times \frac{p_t}{1-p_t},$$
where $|.|$ denotes the number of observations that satisfy a condition, and TP and FP are short for true and false positives, respectively. The probability threshold is denoted as $p_t$, and a subject is treated if the predicted risk $R > p_t$ and untreated otherwise. A decision curve is drawn by plotting the net benefits for a sequence of unique thresholds, and we focus on three five-year ASCVD risk thresholds of 0.25\%, 0.375\%, and 10\%, which is equivalently to the 0.5\%, 0.75\%, and 20\% ten-year probability thresholds recommended by the ASCVD prevention guideline\citep{arnett2019}.

The net benefit is in the unit of true positives, so a value of 5\% indicates that a risk prediction model leads to correctly treating 5 patients per 100 at risk without increasing the number of false interventions. When comparing different risk models, we may also convert the difference in net benefits to the number of additional true positives or false positives that one model can yield or prevent compared to the other, respectively. Let $a = |\mathrm{TP}|/N$, $b=|\mathrm{FP}|/N$, and $k = p_t/(1-p_t)$, then given a threshold, suppose the net benefit for the baseline PCE is $\mathrm{NB}_0 = a_0 - b_0k$ and for a revised model is $\mathrm{NB} = a - bk$, then we have
$$ \mathrm{NB} - \mathrm{NB}_0 = (a-a_0) + (b_0-b)k.$$
Thus, if the revised model achieves the same false positive rate as the baseline PCE, i.e., $b=b_0$, it correctly treats additional $1000(\mathrm{NB} - \mathrm{NB}_0)$ patients per 1000 subjects at risk. Likewise, if the true positive rates are the same, i.e., $a=a_0$, the revised model prevents $b_0-b=1000(\mathrm{NB} - \mathrm{NB}_0)/k$ unnecessary treatment per 1000 subjects at risk.

\section{Results}\label{sect:results}

\subsection{Overall performance}
\begin{figure}[ht!]
\centering
\includegraphics[width=0.8\textwidth]{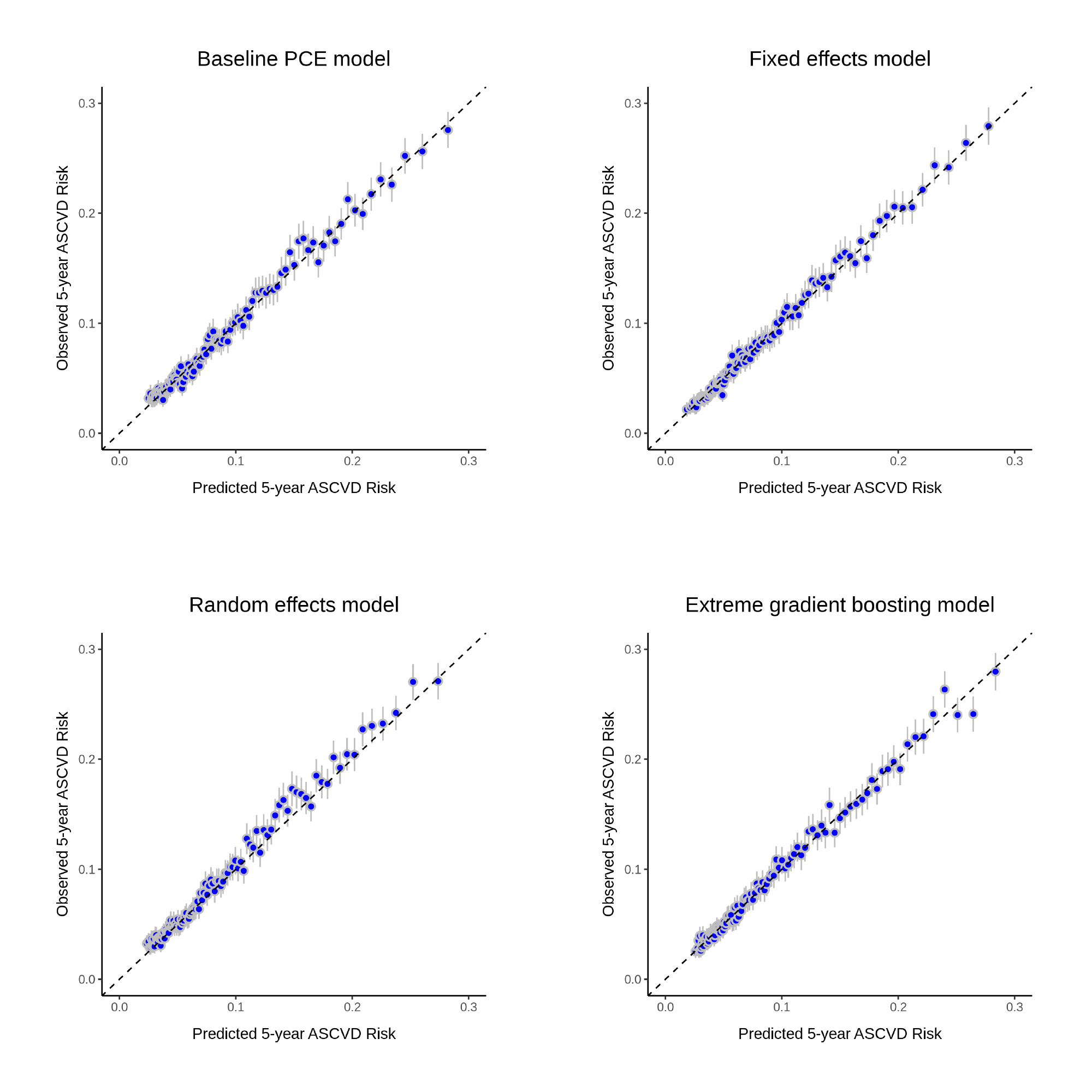}
\caption{Overall calibration plots of baseline PCE and revised models. Observed and estimated risks are compared for 135 risk bins. The fixed effects model and XGB model outperformed the baseline PCE.}\vspace{-1em}
\label{fig:overall_calib}
\end{figure}

Table \ref{t:overall_org_pce} shows the overall performance of the baseline PCE and three revised models. The revised fixed effects model resulted in a slightly higher C-statistic than the baseline PCE. All models showed accurate estimates of the calibration-in-the-large (O/E is close to 1) except that the random effects model overestimated the overall event rate. According to the GND test, the baseline PCE was miscalibrated (p-value $<$ 0.001), but the revised fixed effects and XGB models showed insignificant p-values, which indicate the improvements in calibration. The corresponding calibration plots in Figure \ref{fig:overall_calib} also show that the estimates from the fixed effects model and XGB are closer to the ideal reference line (dotted line) than those from the baseline PCE. Moreover, the calibration slopes estimated using a Poisson model indicate that all models underestimated the five-year ASCVD risk. 

In terms of clinical value, the differences in overall net benefit among the four models and across three risk thresholds are small. If a doctor is willing to prescribe statins to 40 subjects in order to correctly treat one patient who will develop ASCVD within 5 years (threshold = 0.025), the net benefit of baseline PCE and XGB is 0.0736, which means that these two risk estimators are equivalent to a strategy that correctly treats 74 patients per 1000 at risk without increasing the number of false interventions. When the thresholds are 0.0375 and 0.1, the fixed effects model resulted in a slightly higher net benefit than the baseline PCE (difference = 0.004), which maps to correctly treating 4 more patients than the baseline PCE per 1000 patients at risk if the false positive rates are the same, or preventing 103 and 36 unnecessary treatment per 1000 at risk, respectively, if the true positive rates are identical. Figure \ref{fig:overall_nb} shows the decision curves for all risk estimators that capture the net benefits under a range of risk thresholds from 0\% to 10\%. We see that the four risk estimators achieved more clinical benefits than treating everyone in the population when the probability threshold is above 4\%.  

\begin{table}[ht]
\centering
\caption{Overall model performance for estimating the five-year ASCVD risk. \emph{Calib} and \emph{NB} denote calibration and net benefits, respectively. Compared to the baseline PCE, revised risk models improved the calibration performance significantly but only slightly increased the net benefits.}
\label{t:overall_org_pce}
\begin{tabular}{llcccc}
  \hline
Evaluation & Subtype & Baseline PCE & Fixed Effects & Random Effects & XGB \\ 
  \hline 
  C-index & -- & 0.69 (0.69, 0.70) & \textbf{0.70 (0.70, 0.71)} & 0.69 (0.69, 0.70) & 0.69 (0.69, 0.70) \\ \hline
   O/E    & -- & \textbf{1.01 (1.00, 1.02)} & \textbf{1.01 (1.00, 1.02)} & 1.05 (1.04, 1.06) & \textbf{1.01 (1.00, 1.02)} \\ 
  GND & -- & 150.6 ($<$ 0.001) & 113.7 (0.17) & 183.0 ($<$ 0.001) & \textbf{102.5 (0.41)} \\
  \multirow{2}{*}{Calib} & Intercept & \textbf{0.11 (0.10, 0.11)} & 0.12 (0.12, 0.12) & 0.11 (0.11, 0.11) & 0.11 (0.11, 0.11) \\ 
   & Slope & \textbf{1.28 (1.27, 1.29)} & 1.31 (1.30, 1.32) & 1.28 (1.27, 1.29) & 1.29 (1.28, 1.30) \\ \hline
   \multirow{2}{*}{NB} & 2.5\%  & \textbf{0.0736} & 0.0733 & 0.0734 & \textbf{0.0736} \\
                    & 3.75\% & 0.0623 & \textbf{0.0627} & 0.0624 & 0.0625 \\ 
                    & 10\% & 0.0291 & \textbf{0.0295} & 0.0290 & 0.0294 \\  \hline
\end{tabular}
\end{table}

\subsection{Subgroup performance}
\subsubsection{By Comorbidity}
Table \ref{t:sub_org_pce} shows the performance across subgroups defined by the absence or presence of CKD and RA conditions. The baseline PCE showed worse discrimination and calibration-in-the-large, which is as expected as the comorbidity information was not included in the model. In contrast, the revised models that included the CKD and RA covariates slightly improved C-statistic in all subgroups; moreover, they significantly enhanced the calibration in subgroups with small sample sizes (CKD = 1 and RA = 1), as seen in all three calibration metrics. Specifically, the revised models yielded equally accurate estimates of the calibration-in-the-large. Also, the baseline PCE was miscalibrated for all four subgroups per the GND test, while the fixed effects model and the XGB model showed better calibration (P-values $\geq$ 0.05) and better calibration intercepts and slopes in the small-sized subgroups. 

\begin{figure}[ht!]
\centering
\includegraphics[width=0.8\textwidth]{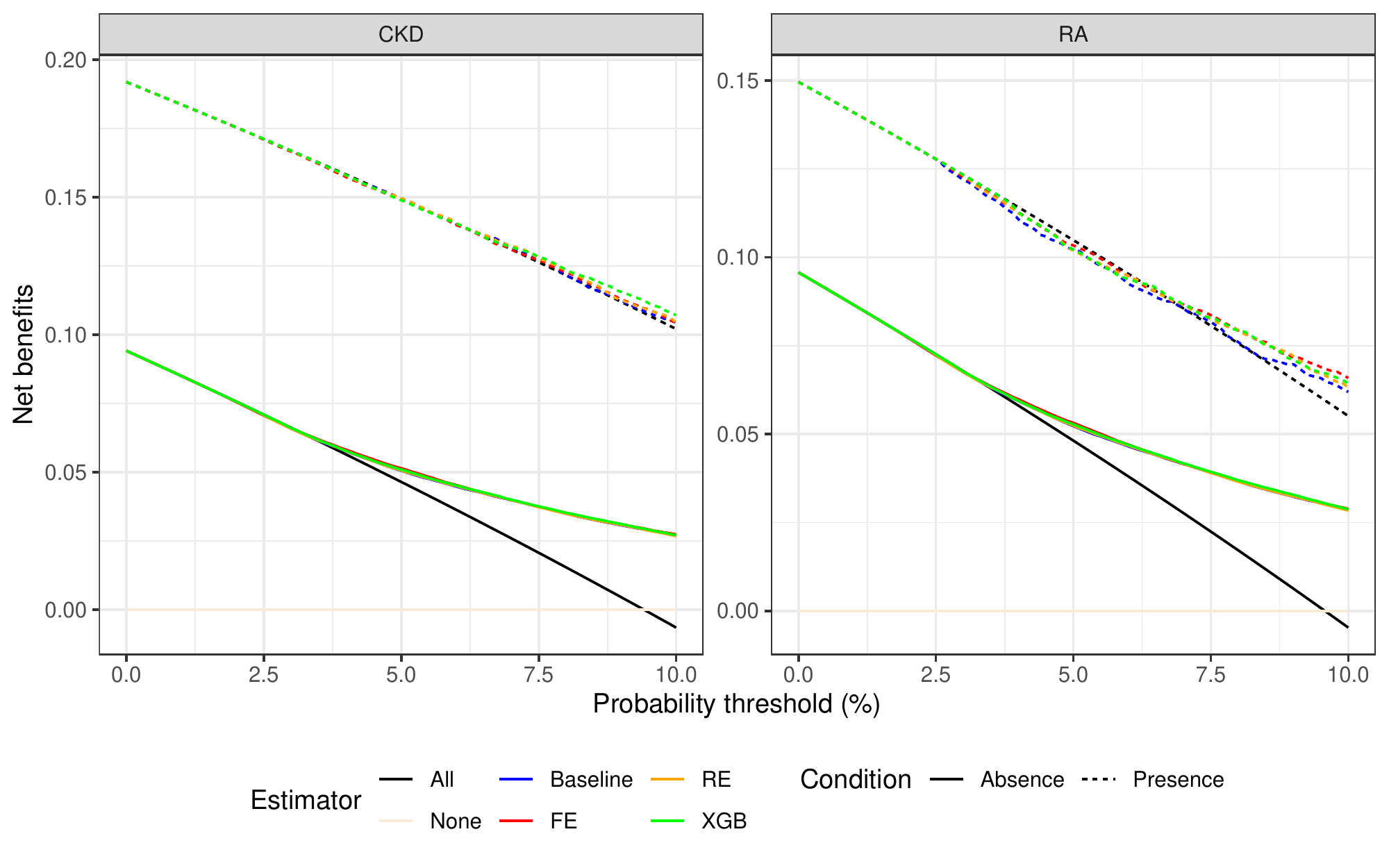}
\caption{Net benefits of risk models in the presence or absence of CKD or RA subgroups. \emph{All} and \emph{None} indicate treating everyone and no one in the cohort, respectively. \emph{Baseline}, \emph{FE}, \emph{RE}, and \emph{XGB} represent the baseline PCE, fixed effects, random effects, and XGB models, respectively. Larger net benefits were shown in the presence of CKD or RA groups, and revised models yielded higher net benefits in the presence of RA subgroup under the risk threshold of 10\%.}\vspace{-1em}
\label{fig:subgroup_nb}
\end{figure}

For the clinical value obtained, Figure \ref{fig:subgroup_nb} shows that the net benefit in the positive comorbidity subgroups are much higher than the negative ones due to higher ASCVD incidence rates in the positive groups. The revised fixed effects model or the XGB model yielded the largest net benefits in all subgroups and under all three risk thresholds, and major improvements occurred in the positive CKD and RA subgroups under the risk threshold of 0.1 (improved net benefits are 0.0027 and 0.0039, respectively), which indicate that they may correctly treating 27 and 39 more patients than the baseline PCE per 1000 at risk if the false positive rates are assumed to be the same, or preventing 243 and 351 unnecessary treatment per 1000 at risk, respectively, if the true positive rates are the same.

\begin{table}
\caption{Model performance across comorbidity subgroups. The baseline PCE is miscalibrated in all subgroups. \emph{Subtype} shows the risk thresholds for the net benefit metric. Revised models significantly improved subgroup model calibration and yielded nontrivially higher net benefits in the positive CKD and RA groups under the threshold of 10\%.}
\label{t:sub_org_pce}
\centering
\resizebox{\textwidth}{!}{\begin{NiceTabular}{lllcccc}[colortbl-like]
  \hline
Evaluation & Subgroup & Subtype & Baseline PCE & Fixed Effects & Random Effects & XGB \\
  \hline
  \multirow{4}{*}{C-index} & CKD = 0 & -- & 0.69 (0.68, 0.70) & \textbf{0.70 (0.69, 0.71)} & 0.69 (0.68, 0.70) & 0.69 (0.69, 0.70) \\ 
                           & \cellcolor{Gray}CKD = 1 & \cellcolor{Gray}-- & \cellcolor{Gray}0.61 (0.58, 0.64) & \cellcolor{Gray}\textbf{0.62 (0.59, 0.65)} & \cellcolor{Gray}0.61 (0.58, 0.64) & \cellcolor{Gray}\textbf{0.62 (0.59, 0.65)} \\ 
                           & RA = 0 & --  & 0.69 (0.69, 0.70) & \textbf{0.70 (0.70, 0.71)} & 0.69 (0.69, 0.70) & 0.70 (0.69, 0.70) \\ 
                           & \cellcolor{Gray}RA = 1 & \cellcolor{Gray}--  & \cellcolor{Gray}0.64 (0.60, 0.68) & \cellcolor{Gray}\textbf{0.65 (0.62, 0.69)} & \cellcolor{Gray}0.64 (0.60, 0.68) & \cellcolor{Gray}0.65 (0.61, 0.68) \\ \hline
  \multirow{4}{*}{O/E} & CKD = 0 & -- & \textbf{1.00 (0.99, 1.01)} & 1.01 (1.00, 1.02) & 1.06 (1.05, 1.07) & 1.01 (1.00, 1.02) \\ 
                       & \cellcolor{Gray}CKD = 1 & \cellcolor{Gray}-- & \cellcolor{Gray}1.09 (1.05, 1.13) & \cellcolor{Gray}\textbf{1.00 (0.96, 1.04)} & \cellcolor{Gray}\textbf{1.00 (0.96, 1.04)} & \cellcolor{Gray}\textbf{1.00 (0.96, 1.04)} \\ 
                       & RA = 0  & -- & \textbf{1.00 (0.99, 1.01)} & 1.01 (1.00, 1.02) & 1.05 (1.04, 1.06) & 1.01 (1.00, 1.02) \\ 
                       & \cellcolor{Gray}RA = 1  & \cellcolor{Gray}-- & \cellcolor{Gray}1.29 (1.22, 1.36) & \cellcolor{Gray}\textbf{1.00 (0.94, 1.05)} & \cellcolor{Gray}1.07 (1.01, 1.13) & \cellcolor{Gray}\textbf{1.00 (0.95, 1.06)} \\ \hline
  \multirow{4}{*}{GND} & CKD = 0 & -- & 161.1 ($<$ 0.001) & 106.5 (\textbf{0.28}) & 161.2 ($<$ 0.001) & 111.3 (0.19) \\ 
                       & \cellcolor{Gray}CKD = 1 & \cellcolor{Gray}-- & \cellcolor{Gray}53.8 (0.005) & \cellcolor{Gray}55.0 (0.004) & \cellcolor{Gray}60.2 ($<$ 0.001) & \cellcolor{Gray}31.2 (\textbf{0.40}) \\ 
                       & RA = 0 & --  & 136.5 (0.009) & 98.7 (\textbf{0.52}) & 164.1 ($<$ 0.001) & 118.5 (0.10) \\ 
                       & \cellcolor{Gray}RA = 1 & \cellcolor{Gray}--  & \cellcolor{Gray}77.3 ($<$ 0.0001) & \cellcolor{Gray}38.6 (0.040) & \cellcolor{Gray}51.0 (0.002) & \cellcolor{Gray}43.6 (0.012) \\ \hline
  \multirow{8}{*}{Calib} & \multirow{2}{*}{CKD = 0} & Intercept & \textbf{0.10 (0.10, 0.11)} & 0.12 (0.11, 0.12) & 0.11 (0.10, 0.11) & 0.11 (0.10, 0.11) \\ 

                         && Slope     & \textbf{1.28 (1.27, 1.29)} & 1.31 (1.30, 1.32) & \textbf{1.28 (1.27, 1.29)} & \textbf{1.28 (1.27, 1.29)} \\ 
                         & \multirow{2}{*}{\centering \cellcolor{Gray}CKD = 1} & \cellcolor{Gray}Intercept & \cellcolor{Gray}0.11 (0.10, 0.12) &\cellcolor{Gray} 0.11 (0.10, 0.13) &\cellcolor{Gray} 0.11 (0.10, 0.12) & \cellcolor{Gray}\textbf{0.10 (0.09, 0.12)} \\ 
                         &  \cellcolor{Gray}  & \cellcolor{Gray}Slope     & \cellcolor{Gray}1.10 (1.07, 1.14) & \cellcolor{Gray}1.12 (1.08, 1.16) & \cellcolor{Gray}1.10 (1.07, 1.14) & \cellcolor{Gray}\textbf{1.10 (1.06, 1.13)} \\ 
                         & \multirow{2}{*}{RA = 0}  & Intercept & \textbf{0.11 (0.10, 0.11)} & 0.12 (0.11, 0.12) & 0.11 (0.11, 0.11) & 0.11 (0.11, 0.11) \\ 
                         &                          & Slope     & \textbf{1.28 (1.27, 1.29)} & 1.31 (1.30, 1.32) & \textbf{1.28 (1.27, 1.29)} & 1.29 (1.28, 1.30) \\ 
                         & \cellcolor{Gray}\multirow{2}{*}{RA = 1}  & \cellcolor{Gray}Intercept & \cellcolor{Gray}0.11 (0.10, 0.13) & \cellcolor{Gray}0.12 (0.10, 0.14) & \cellcolor{Gray}0.11 (0.10, 0.13) & \cellcolor{Gray}\textbf{0.11 (0.09, 0.13)} \\ 
                         &   \cellcolor{Gray}                       & \cellcolor{Gray}Slope     & \cellcolor{Gray}\textbf{1.14 (1.09, 1.20)} & \cellcolor{Gray}1.18 (1.13, 1.23) & \cellcolor{Gray}\textbf{1.14 (1.09, 1.20)} & \cellcolor{Gray}\textbf{1.14 (1.09, 1.20)} \\ \hline
   \multirow{8}{*}{NB} & \multirow{3}{*}{CKD = 0}& 2.5\% & 0.0709 & 0.0706 & 0.0708 & \textbf{0.0710} \\ 
                      &                          & 3.75\% & 0.0596 & \textbf{0.0600} & 0.0597 & 0.0598 \\
                      &                          & 10\% & 0.0270 & \textbf{0.0274} & 0.0269 & 0.0272 \\ 
                      & \multirow{3}{*}{\cellcolor{Gray}CKD = 1} & \cellcolor{Gray}2.5\% &\cellcolor{Gray} 0.1711 & \cellcolor{Gray}0.1709 & \cellcolor{Gray}\textbf{0.1712} & \cellcolor{Gray}\textbf{0.1712} \\ 
                      &         \cellcolor{Gray}                 & \cellcolor{Gray}3.75\% & \cellcolor{Gray}0.1600 & \cellcolor{Gray}0.1598 & \cellcolor{Gray}0.1601 & \cellcolor{Gray}\textbf{0.1603} \\ 
                      &         \cellcolor{Gray}                 & \cellcolor{Gray}10\% & \cellcolor{Gray}0.1044 & \cellcolor{Gray}0.1045 & \cellcolor{Gray}0.1045 & \cellcolor{Gray}\textbf{0.1071} \\
                      & \multirow{3}{*}{RA = 0}  & 2.5\% & 0.0725 & 0.0722 & 0.0724 & \textbf{0.0726} \\ 
                      &                          & 3.75\% & 0.0613 & \textbf{0.0617} & 0.0614 & 0.0614 \\
                      &                          & 10\% & 0.0285 & \textbf{0.0289} & 0.0284 & 0.0288 \\
                      & \cellcolor{Gray}\multirow{3}{*}{RA = 1}  & \cellcolor{Gray}2.5\% & \cellcolor{Gray}0.1278 & \cellcolor{Gray}0.1278 & \cellcolor{Gray}0.1278 & \cellcolor{Gray}0.1278 \\
                      & \cellcolor{Gray}                         & \cellcolor{Gray}3.75\% & \cellcolor{Gray}0.1143 & \cellcolor{Gray}0.1153 & \cellcolor{Gray}0.1153 & \cellcolor{Gray}\textbf{0.1162} \\  
                      & \cellcolor{Gray}                         & \cellcolor{Gray}10\% & \cellcolor{Gray}0.0620 & \cellcolor{Gray}\textbf{0.0659} & \cellcolor{Gray}0.0632 &\cellcolor{Gray} 0.0643 \\  \hline
\end{NiceTabular}}
\end{table}

\subsubsection{By Geographic Location}
\begin{figure}[ht!]
\centering
\includegraphics[width=0.8\textwidth]{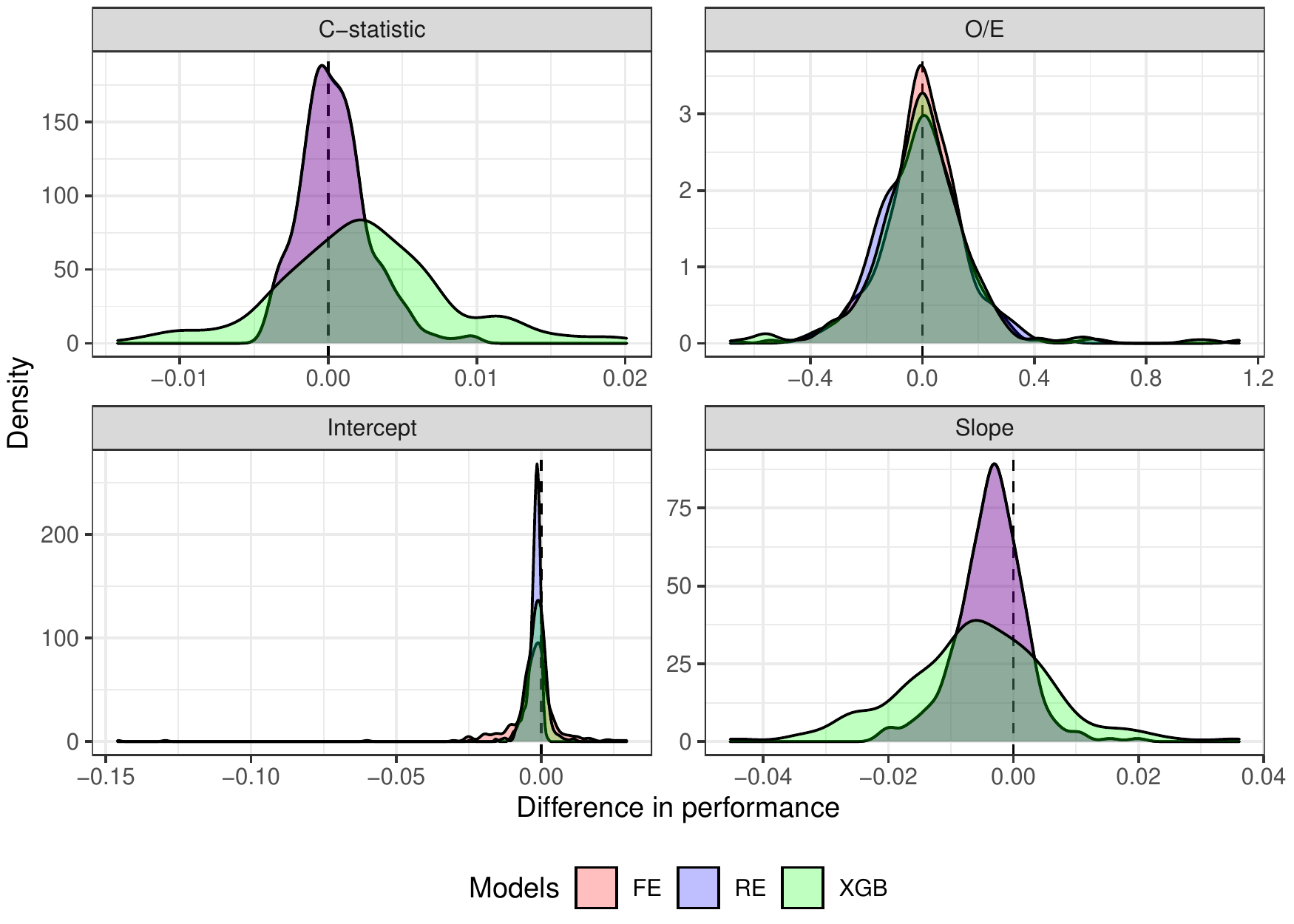}
\caption{Statistical performance of revised models compared to the baseline PCE by residence locations. The XGB model showed higher concordance indexes and better calibration-in-the-large but worse calibration slopes on average.}\vspace{-1em}
\label{fig:sub_zip_other}
\end{figure}

Given the difficulty in tabulating performance of 299 locations, we summarize the statistical performance in Figure \ref{fig:sub_zip_other} as density plots of the difference from baseline for four quantities, the C-statistic, the O/E values, Intercept and Slope. The C-statistic panel shows the density plot of differences in discrimination, calculated as the C-statistic of revised models minus that of the baseline PCE, so a positive value indicates a better discrimination ability than the baseline. The XGB model outperformed the baseline PCE in most of the locations although the magnitude of improvement is small. For overall calibration, the models are compared by how close their O/E values are to the ideal value, i.e., the absolute difference between O/E and 1. The density plot in the O/E panel is skewed to the right, which indicates that revised models resulted in much more accurate O/E estimates than the baseline PCE in some locations. Analogously, the calibration intercept and slope are compared based on the absolute differences between estimates and 0 and 1, respectively, and the revised models underperformed the baseline PCE. 

\begin{figure}[ht!]
\centering
\includegraphics[width=0.7\textwidth]{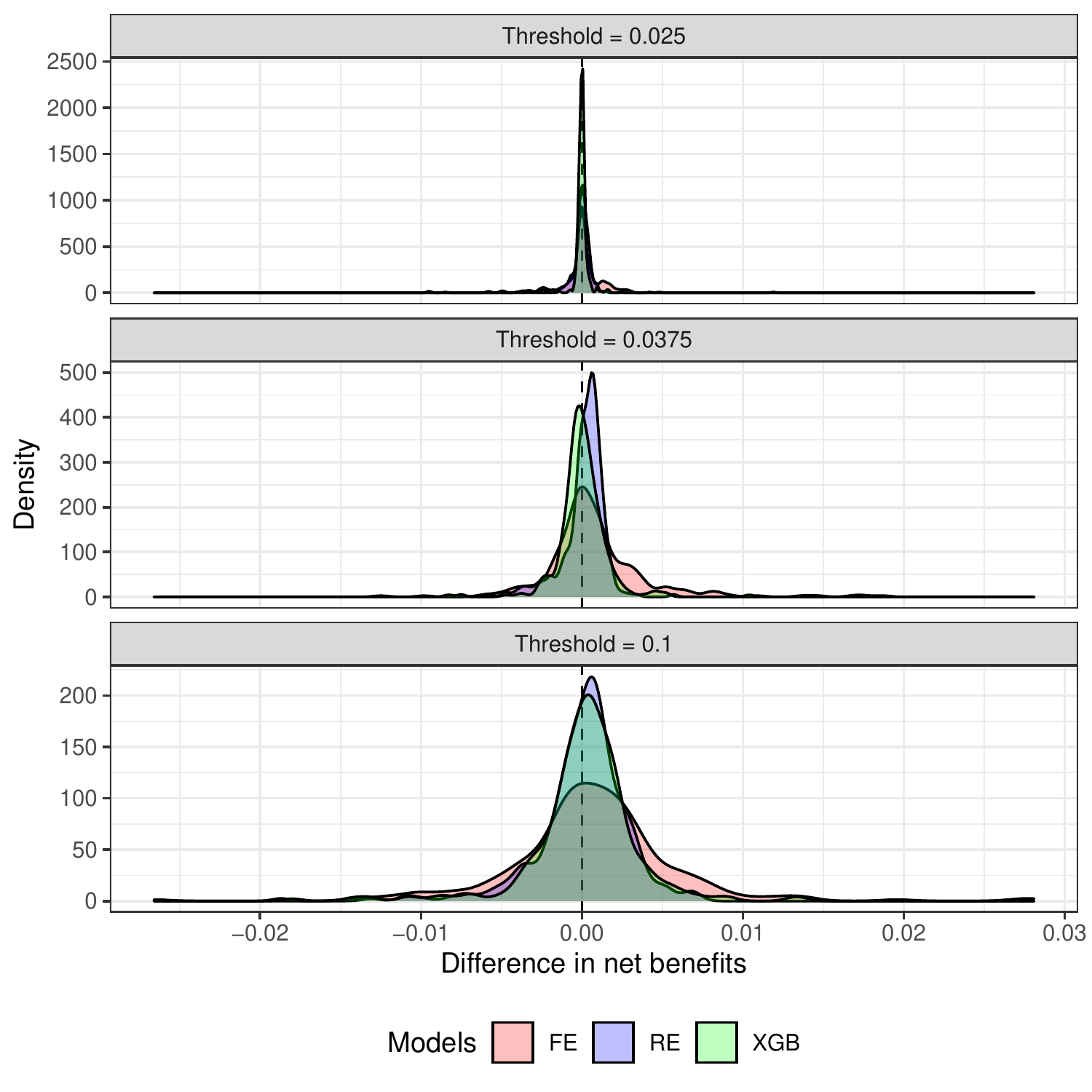}
\caption{Net benefits of revised models compared to the baseline PCE by residence locations. \emph{FE} and \emph{RE} are short for fixed effects and random effects models, respectively. The revised fixed effects and XGB models yielded higher net benefits on average under the risk threshold of 10\%.}\vspace{-1em}
\label{fig:sub_zip_nb}
\end{figure}

\begin{figure}[ht!]
\centering
\includegraphics[width=0.7\textwidth]{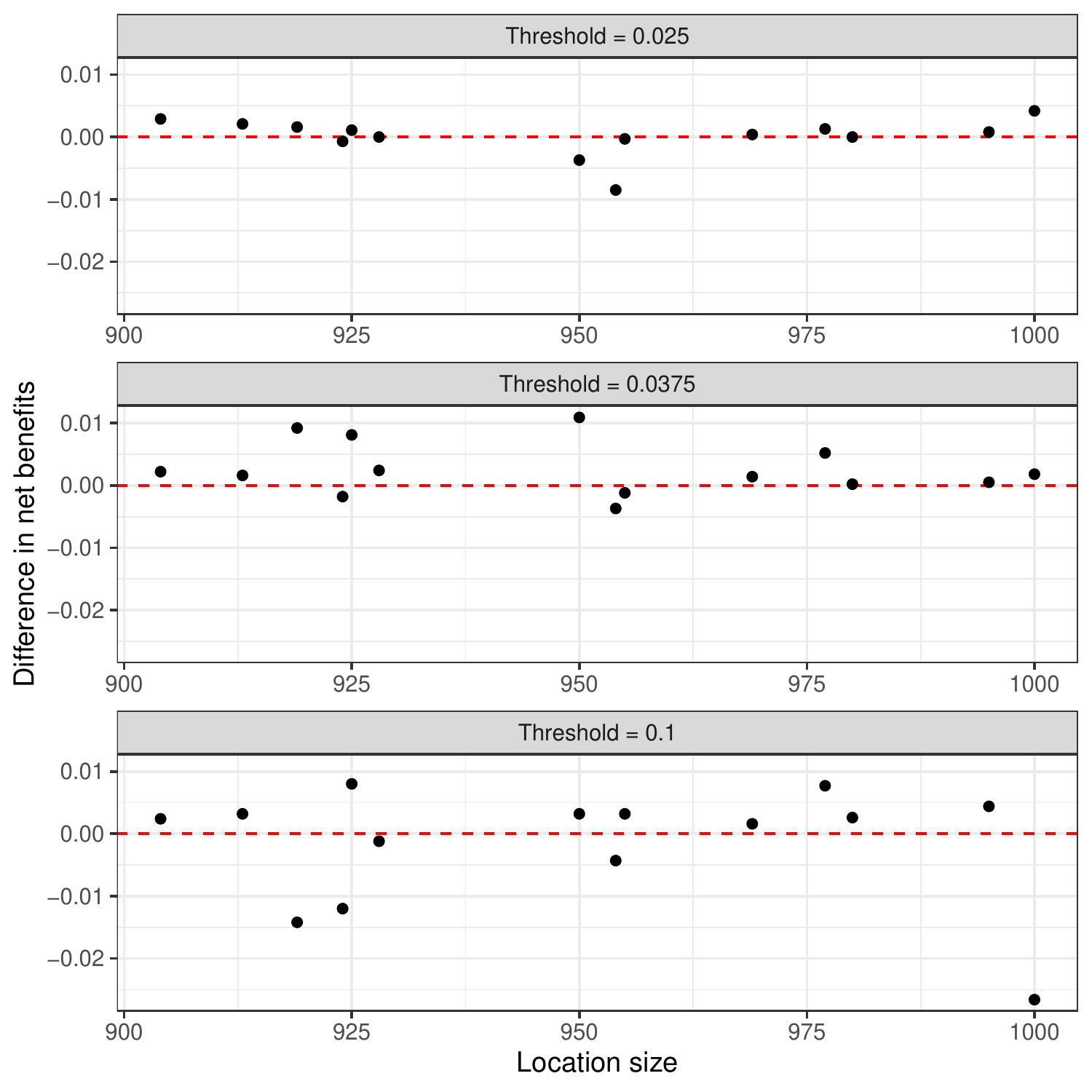}
\caption{Net benefits of revised fixed effects model compared to the baseline PCE in small-sized locations. The fixed effects model resulted in larger net benefits than the baseline PCE in 9 out of 14 locations.}\vspace{-1em}
\label{fig:sub_zip_nb_fe}
\end{figure}

Figure \ref{fig:sub_zip_nb} displays the difference in net benefit of revised models as compared to the baseline PCE cross different local areas. We see the gains in net benefit are tiny when the risk threshold is low (0.025), but they increase as the threshold went up. When the threshold is 0.0375, the random effects model yielded slightly higher net benefit in more locations than the baseline PCE, and when the threshold was increased to 0.1, all three revised models resulted in higher net benefit than the baseline PCE in more than a half of the location subgroups. Furthermore, Figure \ref{fig:sub_zip_nb_fe} shows the difference in net benefit between the fixed effects model and baseline PCE in locations that contain fewer than 1000 subjects. We see the gains in net benefit are positive for 9 out of 14 locations at both the thresholds of 0.0375 and 0.1, which imply that the fixed effects model improved net benefit for areas with small populations. 

\subsection{External validation}
Table \ref{t:overall_starr} shows that the external validation results from comparing the ACC/AHA PCE and the baseline PCE using the STARR data. Without using the race variable and replacing the SBP measure with the hypertension indicator, the baseline PCE achieved a similar discrimination as the ACC/AHA PCE but worse calibration. Moreover, the baseline PCE yielded smaller net benefit than those from the ACC/AHA PCE across all risk thresholds, and smaller than the overall net benefit in the Optum ZIP data (internal validation). 

\section{Discussion}\label{sect:disscusion}
We performed a comprehensive evaluation of the widely used PCE that estimates patients' five-year ASCVD risk and informs treatment initiation with a statin. We re-trained and validated the 2013 ACC/AHA PCE using the large health claims data, Optum ZIP, that includes more than 3.5 million patient records collected from all 50 states in the U.S.. We quantified the clinical utility of risk stratification tools using the net benefit measure obtained from decision curve analysis. We also reported the statistical performance such as the discrimination and calibration abilities of a model. We found that the PCE is miscalibrated especially in the underrepresented subgroups such as patients with CKD or RA conditions and the small-sized local areas. We revised the PCE by incorporating additional comorbidity (CKD and RA) and location (modified zip code) information and applying three different estimation approaches, including fixed effects model, random effects model, and XGB. The revised risk models yielded significant improvements in calibration and net benefit than the baseline PCE in subgroups. 

Incorporating additional information into the baseline PCE improved the model calibration more than discrimination. For both overall and subgroup evaluations, revised models resulted in almost the same C-statistic as the baseline PCE, but the XGB model, for instance, yielded insignificant p-values for the GND tests, which indicate no evidence of miscalibration. Thus, the extra patient-level data on comorbidity and geographic location did not change much of patients' risk ranking but made the estimated risks closer to the observed risks. This finding is encouraging since theoretical results have shown that clinical utility depends on model's calibration\citep{pepe2013net, baker2012evaluating}, and better calibrated models often have larger clinical benefits\citep{van2015calibration}. Our subgroup evaluation also showed consistent results that the calibrated models, e.g., the fixed effects and XGB models, have higher net benefit than the others.  

Compared to the prior work where conclusions are drawn majorly based on statistical performance, we argue that net benefit is a more useful measure because it allows flexible weighing the gain from correctly treating patients who will develop the disease against the loss from assigning unnecessary treatment to subjects who will not develop the disease based on stakeholders' perception. To facilitate interpretation, we converted the values of net benefit into the additional number of patients being correctly treated or the number of unnecessary interventions being prevented per 1000 subjects at risk. This interpretation is especially useful when comparing multiple risk stratification tools. 

According to our results, if the goal is to maximize the expected clinical utility for a population, it may not be necessary to revise the baseline PCE due to the trivial gains. However, the revised models yielded larger net benefit than the baseline PCE in patients with CKD or RA when the five-year ASCVD risk threshold is higher than or equal to 10\%, which imply that model revision may be useful in high-risk subgroups; besides, we found that model revision is also helpful for improving the net benefit in locations with small populations.

We also observed large heterogeneity in model performance across subgroups. The discrimination power of all four models are lower in patients with comorbidity whose event rates are nearly twice as high as those without comorbidity. Such differential performance across subgroups is referred to as \emph{metric disparity} in model fairness assessment, and is necessary to address to mitigate health inequalities\citep{pfohl2022net}. As expected, such metric disparity is exacerbated when the number of subgroups grows larger, such as across 299 geographic locations. Prior works attempted to diminish such fairness violation by incorporating fairness constraints into a training objective function but resulted in worse statistical performance and lower clinical utilities. So, although health equality is highly desired, how to achieve this goal without losing other valuable proprieties of a risk model is still an open research question.  

Our evaluation of the ASCVD risk models has the following strengths: First, we utilized the real-world and large claims data, Optum ZIP, that contains subject-level information on comorbidities and residence zip code. The latter enabled us to evaluate model performance across a large number of geographic locations with confidence in the sense that each location subgroup contains at least 900 patients in the testing data. Second, our revised models incorporate risk factors beyond the ones already included in the ACC/AHA PCE, and our evaluations were performed in the novel subgroups defined by these additional covariates rather than subgroups defined by race and gender. So, our study provides new insights on differential performance by subgroup. Third, we focused on comparing models' clinical utility under different thresholds, so our results directly inform the gains obtained from deploying revised models to guide interventions. Note that the chosen threshold reflects decision-makers' perception of the benefits of intervening on a true case over the harms of unneeded treatment, so it may be very different for different stakeholders.  

Our study has several limitations. First, the lack of SBP and race information in the Optum ZIP data. This weakness forced us to use a surrogate which underperformed the 2013 ACC/AHA PCE. Despite this limitation, our comparison of revised models to the baseline PCE is valid since all models were trained using the Optum ZIP data with the same drawback. Second, we combined some of the original locations defined by 5-digits zip code to avoid subgroups with extremely small sizes. Even though the characteristics within each merged location may become more heterogeneous and less clear, merging small areas was necessary to have sufficient sample sizes that are critical for conducting reliable subgroup evaluations. As we combined each original location with its nearest neighbours iteratively until the minimum sample size of 3000 is reached, the correlation within each original location subgroup has been maximally reserved. Third, due to computation burden, we used a relatively simple discrimination metric that assumes noninformative censoring. However, we examined that using the IPCW C-statistic that allows modeling censoring mechanism as a function of covariate adjustment yielded very similar results.

We focused on the ASCVD risk stratification tools in this work, but our model revision and evaluation strategies can be easily applied to many other risk calculators such as CHA\textsubscript{2}DS\textsubscript{2}VASc score for atrial fibrillation stroke and the Model for End-stage Liver Disease (MELD) score to quantify the gains from learning new models. We appreciate the large amount of prior efforts on refining clinical risk calculators (e.g., the ones included in MDcalc), but we believe that future work needs to examine model improvements in terms of the clinical utility and consequences. Our study can be viewed as such an example where we judge whether or not a revised model is worth being deployed based on its clinical value, that is, its consequences on guiding interventions. 

\section{Conclusion}
We found that revising the PCE using additional comorbidity and location information significantly enhanced the overall and subgroup calibration. However, such improvements did not always translate into gains in clinical utility measured in terms of net benefit. Thus, we recommend that future efforts on improving clinical risk calculators should include evaluation metrics that consider clinical consequences such as net benefits. 

\section*{Acknowledgements}
This work was supported by R01 HL144555 from the National Heart, Lung, and Blood Institute (NHLBI). We appreciate the great help from Dr. Stephen Pfohl on data extraction and reviewing manuscript. 

\bibliographystyle{unsrtnat} 
\bibliography{references}

\newpage 
\section*{Legends of Figures}
Figure \ref{fig:overall_calib}: Overall calibration plots of baseline PCE and revised models. Observed and estimated risks are compared for 135 risk bins. The fixed effects model and XGB model outperformed the baseline PCE.

Figure \ref{fig:subgroup_nb}: Net benefits of risk models in the presence or absence of CKD or RA subgroups. \emph{All} and \emph{None} indicate treating everyone and no one in the cohort, respectively. \emph{Baseline}, \emph{FE}, \emph{RE}, and \emph{XGB} represent the baseline PCE, fixed effects, random effects, and XGB models, respectively. Larger net benefits were shown in the presence of CKD or RA groups, and revised models yielded higher net benefits in the presence of RA subgroup under the risk threshold of 10\%.

Figure \ref{fig:sub_zip_other}: Statistical performance of revised models compared to the baseline PCE by residence locations. The XGB model showed higher concordance indexes and better calibration-in-the-large but worse calibration slopes on average.

Figure \ref{fig:sub_zip_nb}: Net benefits of revised models compared to the baseline PCE by residence locations. \emph{FE} and \emph{RE} are short for fixed effects and random effects models, respectively. The revised fixed effects and XGB models yielded higher net benefits on average under the risk threshold of 10\%.

Figure \ref{fig:sub_zip_nb_fe}: Net benefits of revised fixed effects model compared to the baseline PCE in small-sized locations. The fixed effects model resulted in larger net benefits than the baseline PCE in 9 out of 14 locations.

\end{document}


\textbf{Title}: Clinical Utility Gains from Incorporating Comorbidity and Geographic Location Information into Risk Estimation Equations for Atherosclerotic Cardiovascular Disease 

\vspace{0.25in}

\textbf{Corresponding author}: Yizhe Xu\\ 
\textbf{E-mail}: yizhex@stanford.edu \\
\textbf{Postal address}: 3180 Porter Dr., Palo Alto, CA, 94304\\
\textbf{Telephone}: (801) 433-7346

\vspace{0.25in}

\textbf{Co-authors}: Yizhe Xu, Agata Foryciarz, Ethan Steinberg, and Nigam H. Shah\\
\textbf{Affiliation}: (shared by all co-authors) Stanford Center for Biomedical Informatics Research, Stanford University, Stanford, CA.
\vspace{0.25in}

\textbf{Keywords}: Atherosclerotic Cardiovascular Disease, Clinical Utility, Model Calibration, Net Benefit, Pooled Cohort Equations, Subgroup Performance

\vspace{0.25in}
\textbf{Word count}: 6,000

\newpage

\beginsupplement

\section{Appendix A: Additional details of the applied models}
\subsection{Extreme Gradient Boosting (XGB)}\label{XGB}
As of gradient boosting machine \citep{Friedman2001}, XGB is an ensemble tree method that learns the mapping of the outcome surface as a function of baseline covariates $F(x; \theta)$ via a greedy stagewise algorithm. Specifically, an XGB model is an additive expansion of sequentially fitted trees, i.e., $F(X) = \sum_{m=1}^{M} w_m h(X, \beta_m)$, where $M$ denotes the total number of trees and $w_m$ is the weight of the tree in stage $m$. Let $(t_i, \delta_i)$ be the event time and event indicator, respectively, of a survival outcome, XGB constructs the weak learner $h(X, \beta_m)$ as:
$$\beta_m = \argmin_\beta\sum_{i=1}^{n}(z_i - h(X_i, \beta_m))^2, \,\,\, \mathrm{where} \,\, z_i = \delta_i - \sum_{j:\,t_i\geq t_j}\delta_j \frac{e^{F(X_i)}}{\sum_{k:\,t_k\geq t_j}e^{F(X_k)}},$$ 
and $z_i$ is referred to as \emph{negative gradient} or \emph{working response}. Then, XGB iteratively fits a non-linear PH model to the outcome $(t_i, \delta_i)$ with an offset $F_{m-1}(X_i)$, the model fitting from last stage, and a predictor $h(X, \beta_m)$, so
$w_m$ is the coefficient estimated by minimizing the current residual loss:
$$w_m = \argmin_w\sum_{i=1}^{n} \mathcal{L}(t_i, \delta_i, F_{m-1}(X_i) + w h(X, \beta_m)), $$
where $\mathcal{L}(\cdot)$ is the negative Cox's log-partial likelihood loss that accounts for right censoring \citep{Ridgeway1999}: 
$$\mathcal{L}(F|t_i, \delta_i, X_i) = -\sum_{i=1}^{n} \delta_i\left\{F(X_i)-\mathrm{log}\left(\sum_{j:\,t_j\geq t_i}e^{F(X_j)}\right)\right\}.$$
With an initial $\what{F}_0(X_i) = 0$ and the learned weights $w_m$, XGB updates the function approximator $F(X_i)$ for the next stage as:
$$\what{F}_{m}(X_i) = \what{F}_{m-1}(X_i) + w_m h(X, \beta_m), \,\, m = 1, ..., M.$$

XGB applies shrinkage after each step tree boosting (multiple $w_m$ by a factor $\nu$), row subsampling (a fraction of the training data randomly selected to propose the next tree in expansion), and column subsampling (a random sample of the features in training data are used for tree construction) to control overfitting.

\subsection{IPCW C-statistic}\label{IPCW-C}
In contrast to Harrell's C-statistic \citep{harrell1996multivariable}, the inverse-probability-censoring weighted (IPCW) C-statistic is an alternative discrimination metric that accounts for the unusable pairs. Specifically, the IPCW C-statistic utilizes a working model to estimate the survival probability of getting censored $\what{W}_{ij}$ as a function of covariates then re-weight $\what{C}_\mathrm{HC}(t)$ for censored data as follows  \citep{gerds2013estimating}:
$$\what{C}_\mathrm{IPCW}(t) = \frac{\frac{1}{m^2}\sum_{i=1}^{m}\sum_{j=1}^{m}\ind{M_n(t, X_i)>M_n(t, X_j)} \mathbb{1}\{\widetilde{T}_i < \widetilde{T}_j\} \widetilde{N}_i(t) \what{W}_{ij}^{-1}}{\frac{1}{m^2}\sum_{i=1}^{m}\sum_{j=1}^{m}\ind{\widetilde{T}_i < \widetilde{T}_j} \widetilde{N}_i(t) \what{W}_{ij}^{-1}},$$
The IPCW C-statistic is a more advanced metric that accounts for scenarios where the censoring distribution depends on baseline covariates; however, its implementation via R package \emph{pec} on large data sets is time-consuming, especially when the inference is of interest. We eventually choose to apply the Harrell's C-statistic as it yielded very similar results as the IPCW C-statistic under several estimation cases

\section{Appendix B: Additional analysis results}
\begin{table}[ht]
\centering
\caption{Concept and concept codes used to extract the Optum and STARR cohorts.}
\label{t:concept_code}
\begin{tabular}{ll}
  \hline
 Concept                  & Code \\\hline
 Cardiovascular disease (ICD-9)&  410*, 411*, 413*, 414*, 430*, 431*, \\
 & 432*, 433*, 434*, 436*, 427.31, 428*\\
 Coronary heart disease (ICD-9)& 411*, 413*, 414*\\
 Myocardial infarction (ICD-9)& 410*\\
 Stroke (ICD-9)& 430*, 431*, 432*, 433* (exclude 433.*0), \\
 & 434* (exclude 434.*0), 436*\\
 Hypertension (ICD-9) & 401*, 402*, 403*, 404*, 405*\\
 Type 1 diabetes (SNOMED) & 46635009, 444073006, 420868002\\
 Type 2 diabetes (SNOMED) & 74627003, 268519009, 73211009\\
 Chronic kidney disease (SNOMED) & 42399005, 28689008, 36171008, 709044004, 197661001,\\
 & 38481006, 52254009, 52845002, 127013003, 36184004 \\
 & (exclude 50581000, 733839001, 102455002, 43629001, \\
 & 198949009, 200118004, 707742001, 14669001, 609452007\\
 Rheumatoid arthritis (SNOMED) & 69896004\\
 Statin (ATC) & C10AA0[1-8]\\
 Systolic blood pressure (LONIC) & 76534-7, 8479-8, 8480-6\\
 High-density lipoprotein (LONIC) & 12772-0, 54372-8, 2085-9, 27340-9, 14646-4, \\	
& 12771-2, 53135-0, 18263-4, 50840-8, 49130-8\\
 Total cholesterol (LONIC)& 12183-0, 14647-2, 2093-3\\\hline
\end{tabular}
\end{table}

\begin{table}[ht]
\centering
\caption{Overall performance of the 2013 ACC/AHA PCE and the re-trained baseline PCE in STARR. The surrogate baseline PCE resulted in similar discrimination performance as the ACC/AHA PCE but worse calibration and net benefits.}
\label{t:overall_starr}
\begin{tabular}{lcccc}
  \hline
                   & 2013 ACC/AHA PCE &  Re-trained Baseline PCE \\\hline
 C-index & 0.69 (0.66, 0.72) & 0.70 (0.67, 0.73) \\\hline
 GND & 83 ($<$0.0001) & 3827 ($<$0.0001) \\\hline
\multicolumn{3}{l}{Net benefit}\\
\,\,\,2.5\% & 0.0227 & 0.0181 \\ 
\,\,\,3.75\% & 0.0168 & 0.0058 \\ 
\,\,\,10\% & 0.0022 & -0.0208 \\ 
\hline
\end{tabular}
\end{table}

\begin{table}[ht]
\centering
\caption{Discrimination of the baseline PCE evaluated by two different concordance metrics. Applying the IPCW C-statistic where censoring is modeled as a function of covariates \citep{gerds2013estimating} yielded very similar results as the Harrell's C-statistic that assumes noninformative censoring \citep{pencina2004overall}.}
\label{t:cindex_comp}
\begin{tabular}{lcc}
  \hline
         & Harrell's C-statistic    & IPCW C-statistic \\\hline
 Overall & 0.69 (0.69, 0.70) & 0.69 \\\hline
 CKD=1   & 0.61 (0.58, 0.64) & 0.62 \\\hline
 CKD=0   & 0.69 (0.68, 0.70) & 0.69 \\
\hline
\end{tabular}
\end{table}

\begin{figure}[ht!]
\centering
\includegraphics[width=0.6\textwidth]{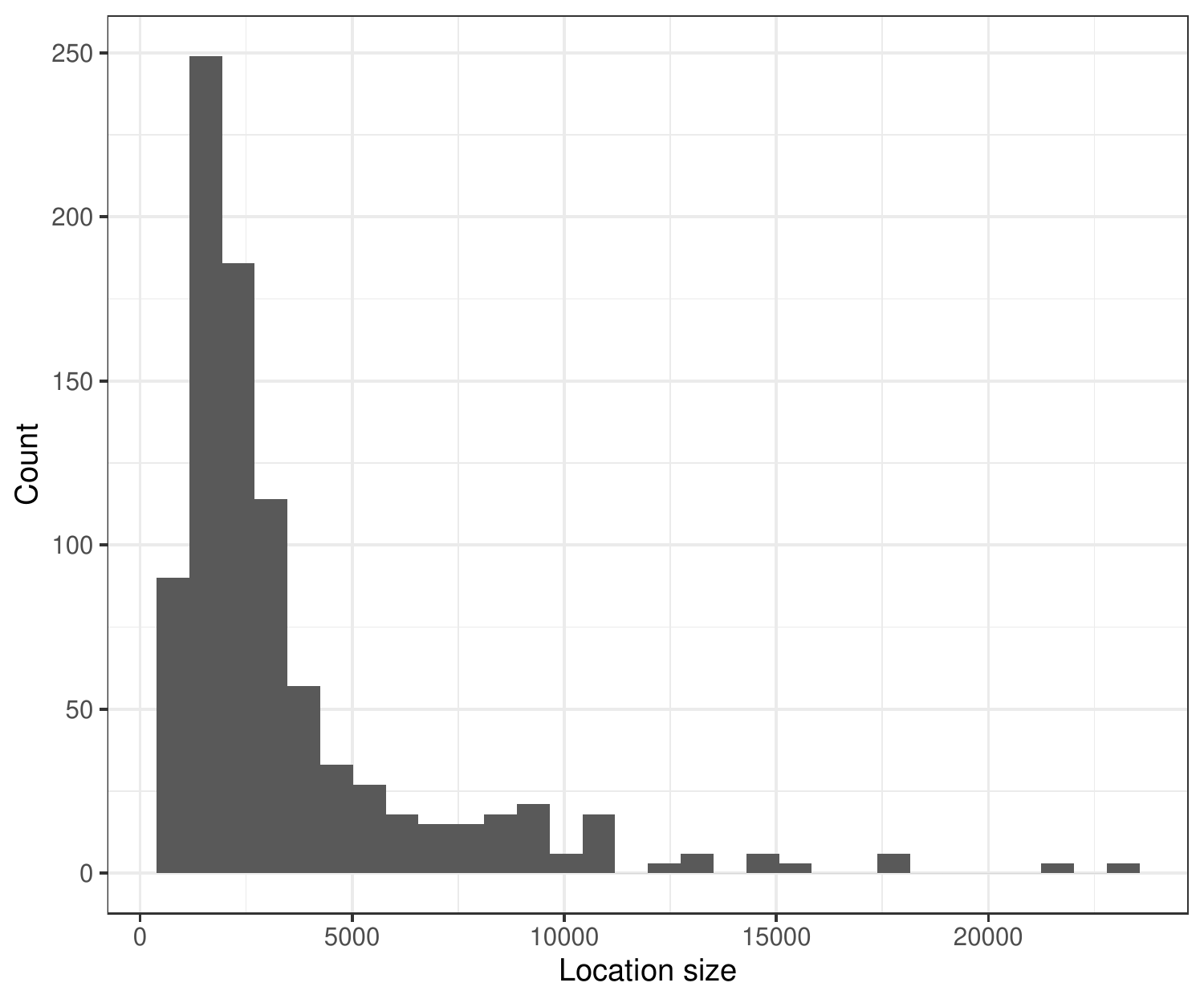}
\caption{Distribution of sample sizes across 299 residence locations.}\vspace{-1em}
\label{fig:hist_local}
\end{figure}

\begin{figure}[ht!]
\centering
\includegraphics[width=0.7\textwidth]{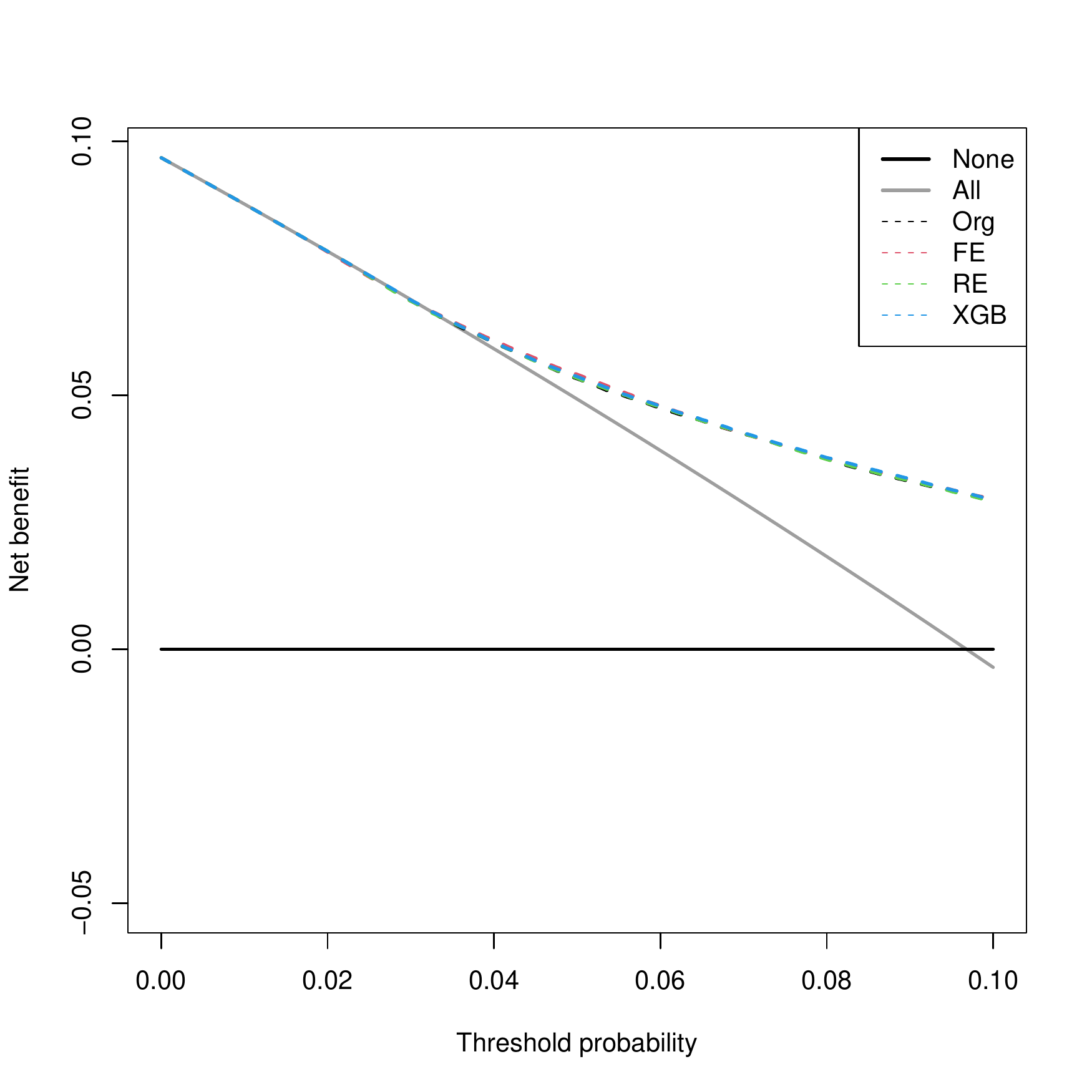}
\caption{Decision curves under Different Risk Estimators. Overall, the revised models resulted in similar clinical utilities as the baseline PCE.}\vspace{-1em}
\label{fig:overall_nb}
\end{figure}

\begin{figure}[ht!]
\centering
\includegraphics[width=0.9\textwidth]{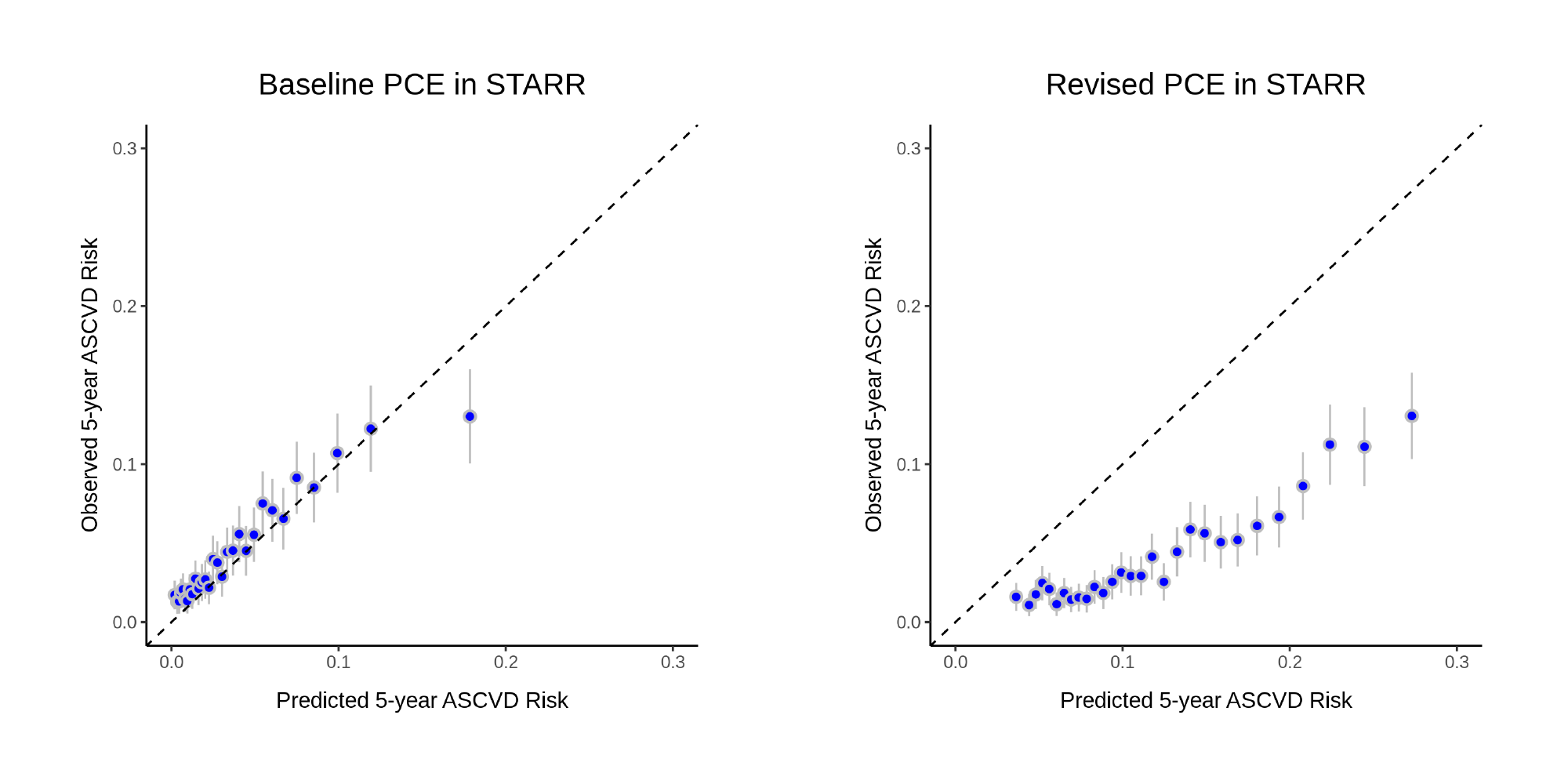}
\caption{Calibration Plots of the 2013 ACC/AHA PCE and baseline PCE in STARR. The baseline PCE was poorly calibrated and overestimated the five-year ASCVD risks.}\vspace{-1em}
\label{fig:overall_calib_starr}
\end{figure}

\vspace{10pt}
\newpage
\bibliographystyle{unsrtnat} 
\bibliography{references}